\documentclass[9pt]{article}

\usepackage[a4paper,total={7in, 10in}]{geometry}
\usepackage{hyperref}
\usepackage{dcolumn}

\usepackage[utf8]{inputenc}
\usepackage[T1]{fontenc}
\usepackage{textcomp}
\usepackage{newtxmath}

\usepackage{authblk}
\usepackage{mymacros}

\usepackage{algorithm,algpseudocode}\usepackage{bm}

\usepackage{graphicx}
\usepackage{cleveref}
\usepackage[caption=false]{subfig}
\usepackage{tabularx}
\usepackage{mathtools}

\usepackage{threeparttable}
\usepackage{multirow}
\usepackage{booktabs}
\usepackage[style=numeric,backend=biber,url=false,sorting=none]{biblatex}
\usepackage{siunitx}[=v2]
\usepackage{physics}
\usepackage{enumitem}

\usepackage[scr=boondoxo]{mathalpha}

\usepackage{xcolor}

\newcommand{\singlespacesmall}{\linespread{1}\small}

\DeclareRobustCommand{\rchi}{{\mathpalette\irchi\relax}}
\newcommand{\irchi}[2]{\raisebox{\depth}{$#1\chi$}}

\newcommand\plane[2]{$#1$\nobreakdash--$#2$~plane}

\newcommand{\OO}{\ensuremath{\mathscr{O}}}
\newcommand{\PP}{\ensuremath{\mathscr{P}}}
\newcommand{\oo}{\ensuremath{\mathscr{o}}}

\newcommand{\FF}{\bm{\mathcal{F}}}
\newcommand{\SSn}{\bm{\mathcal{S}}}

\DeclareMathOperator*{\argmin}{argmin}

\newcommand{\mat}[1]{\mathcal{#1}}
\newcommand{\vect}[1]{\bm{#1}}

\newcommand{\yy}{\vect{y}}

\newcommand{\bb}{\vect{b}}

\newcommand{\uu}{\vect{u}}
\newcommand{\vectG}{\vect{G}}
\newcommand{\vectI}{\vect{I}}
\newcommand{\vectU}{\vect{U}}
\newcommand{\vectm}{\vect{m}}
\newcommand{\vectphi}{\vect{\phi}}
\newcommand{\vectchi}{\vect{\rchi}}
\newcommand{\vectalpha}{\vect{\alpha}}
\newcommand{\vecte}{\vect{e}}
\newcommand{\vectUstar}{\vect{U}_\star}

\newcommand{\xx}{\vect{x}}
\newcommand{\rr}{\vect{r}}
\newcommand{\kk}{\vect{k}}

\newcommand{\GG}{\mat{G}}
\newcommand{\matR}{\mat{R}}

\newcommand{\pgrad}{\boldsymbol{\partial}}

\newcommand{\Rel}[1]{\mathfrak{R}[#1]}
\newcommand{\Img}[1]{\mathfrak{I}[#1]}
\newcommand{\Reals}{\mathbb{R}}

\renewcommand{\diag}[1]{\operatorname{diag}\left(#1\right)}

\newcommand{\ie}{\emph{i.e.},}

\newcommand{\etal}{\emph{et al}}

\newcommand{\viz}{\textit{viz.},}

\algnewcommand\algorithmicinit{\textbf{Initialize:}}
\algnewcommand\Initialize{\item[\algorithmicinit]}

\newcounter{parentalgorithm}



\begin{document}
\title{Imaging extended single crystal lattice distortion fields with multi-peak Bragg ptychography}

\author[1]{S. Kandel}
\author[2,*]{S. Maddali}
\author[3]{M. Allain}
\author[4]{X. Huang}
\author[5,**]{Y. S. G. Nashed}
\author[6]{C. Jacobsen}
\author[2,$\dagger$]{S. O. Hruszkewycz}

\affil[1]{X-ray Science Division, Argonne National Laboratory, Lemont IL 60439, USA}
\affil[2]{Materials Science Division, Argonne National Laboratory, Lemont IL 60439, USA}
\affil[3]{Aix-Marseille Univ, CNRS, Centrale Marseille, Institut Fresnel, Marseille, France}
\affil[4]{National Synchrotron Light Source II, Brookhaven National Laboratory, Upton, NY, 11973, USA}
\affil[5]{Mathematics \& Computer Science Division, Argonne National Laboratory, Lemont IL 60439, USA}
\affil[6]{Department of Physics and Astronomy, Northwestern University, Evanston, IL 60208, USA}
\affil[*]{Currently at KLA Corporation, CA (USA)}
\affil[**]{Currently at Bristol Myers Squibb, NJ (USA)}
\affil[$\dagger$]{Corresponding author: \texttt{shrus@anl.gov}}

\date{}

\maketitle
\begin{abstract}
	We describe a phase-retrieval-based imaging method to directly spatially resolve the vector lattice distortions in an extended crystalline sample by explicit coupling of independent Bragg ptychography data sets into the reconstruction process. 
Our method addresses this multi-peak Bragg ptychography (MPBP) inverse problem by explicit gradient descent optimization of an objective function based on modeling of the probe-lattice interaction, along with corrective steps to address spurious reconstruction artifacts. Robust convergence of the optimization process is ensured by computing exact gradients with the automatic differentiation capabilities of high-performance computing software packages.  
We demonstrate MPBP reconstruction with simulated ptychography data mimicking diffraction from a single crystal membrane containing heterogeneities that manifest as phase discontinuities in the diffracted wave. 
We show the superior ability of such an optimization-based approach in removing reconstruction artifacts compared to existing phase retrieval and lattice distortion reconstruction approaches. 
 \end{abstract}

\section{Introduction}
\label{sec:intro}
Bragg coherent diffraction imaging (BCDI) is a valuable tool to spatially resolve the lattice strain in crystalline materials at the scale of 
\SI{\sim 10}{\nm} by using coherent x-rays as a probe~\cite{robinson_nmat_2009,RobinsonMRS2004,robinson_applss_2001}. 
Bragg ptychography, the scanning variant of BCDI, is used for extended crystalline samples larger than the beam size~\cite{godard_natcomm_2011,hruszkewycz_nl_2012,hruszkewycz_nm_2017}. 
Both techniques have been applied to a variety of materials science problems, such as the performance of  batteries~\cite{ulvestad_nl_2014,ulvestad_science_2015,singer_natenergy_2018}, catalytic phenomena at metal interfaces~\cite{komanicky_electrochimica_acta_2013,cha_adv_func_mat_2017,kawaguchi_prl_2019,kawaguchi_jkps_2019}, and characterization of quantum sensors~\cite{hruszkewycz_prm_2018}. These conventional BCDI methods enable the spatial resolution of a single component of the three-component elastic lattice distortion field. 
This component corresponds to the Bragg diffraction vector of momentum transfer that comes about via a specific orientation of the single crystal lattice of the sample with respect to the beam (one of the crystal's Bragg conditions). Reorienting the crystal into other configurations with respect to the incident x-ray probe permits the independent components of the lattice distortion to be imaged.
When three or more independent Bragg conditions can be satisfied and measured with BCDI, it is possible to unambiguously resolve the lattice displacement vector field within the sample. 

Recent efforts to simultaneously reconstruct all components of internal lattice distortion fields with BCDI have shown promising results with simulations~\cite{newton_prb_2020,gao_prb_2021} as well as experiments~\cite{hofmann_prm_2020,wilkin_prb_2021}
The experimental works have employed different combinations of well-known fixed point iterative projection algorithms~\cite{marchesini_rsi_2007}, typically in parallel reconstruction threads. 
The solutions of these parallel reconstructions are analyzed for desirable properties in order to seed the next generation of randomized solutions (\emph{i.e.}, the `guided' approach using genetic algorithms~\cite{ulvestad_sr_2017}). 
All the vector field reconstruction demonstrations referenced above explicitly minimize the discrepancy between the bulk distortion field and the computed phase from conventional phase retrieval. 
A recent alternate approach seeks to directly and simultaneously minimize the discrepancy in the measured and estimated coherent diffraction signals in an all-encompassing gradient descent optimization scheme ~\cite{maddali_npj_compmat_2023}.  
However, this specific approach pertains only to full coherent illumination of compact crystals, and the method is yet to be demonstrated on the relatively data-intensive Bragg ptychography problem with extended single-crystal samples.

We address the problem of multi-peak Bragg ptychography (MPBP) in this paper through simulations of independent ptychographic data sets and subsequent reconstruction of a two-dimensional projection of the underlying crystal lattice fields; this extends upon the preliminary work first recorded in \cite{kandel_diss_2021}.
The MPBP measurement approach  follows that of Bragg projection ptychography \cite{hruszkewycz_nl_2012,hruszkewycz_prl_2013,hruszkewycz_pra_2016} where, at a given angle that fulfills the Bragg condition, data is collected at multiple overlapping beam positions. 
In MPBP this process is repeated at different Bragg conditions.    
In terms of image reconstruction, MPBP employs a joint optimization scheme similar to ~\cite{maddali_npj_compmat_2023} for all the the measured ptychographic data sets corresponding to the independent Bragg conditions.  
Methods such as MPBP are poised to take advantage of the orders of magnitude increases in coherent flux at hard x-ray energies afforded by fourth-generation synchrotron light sources to image local displacement fields in crystals. 

This paper is organized in the following manner: 
Section~\ref{sec:fmodel} introduces the MPBP forward model connecting the lattice distortion to the measured coherent diffraction signal for a single crystal. 
Section~\ref{sec:optimization} lays out specifics of the numerical optimization problem, including the gradient descent scheme, the expected phase-wrap degeneracies in the solution and their resolution. 
Sections~\ref{sec:numeric} and~\ref{sec:results} describe the parameters of the numerical demonstrations in this paper, and the reconstruction results respectively. 
Sections~\ref{sec:discussion} closes with a discussion of the results and concluding remarks.

\section{Forward model for multi-peak Bragg ptychography (MPBP)}
\label{sec:fmodel}
\begin{figure}[th]
	\centering
	\includegraphics[width=0.9\linewidth]{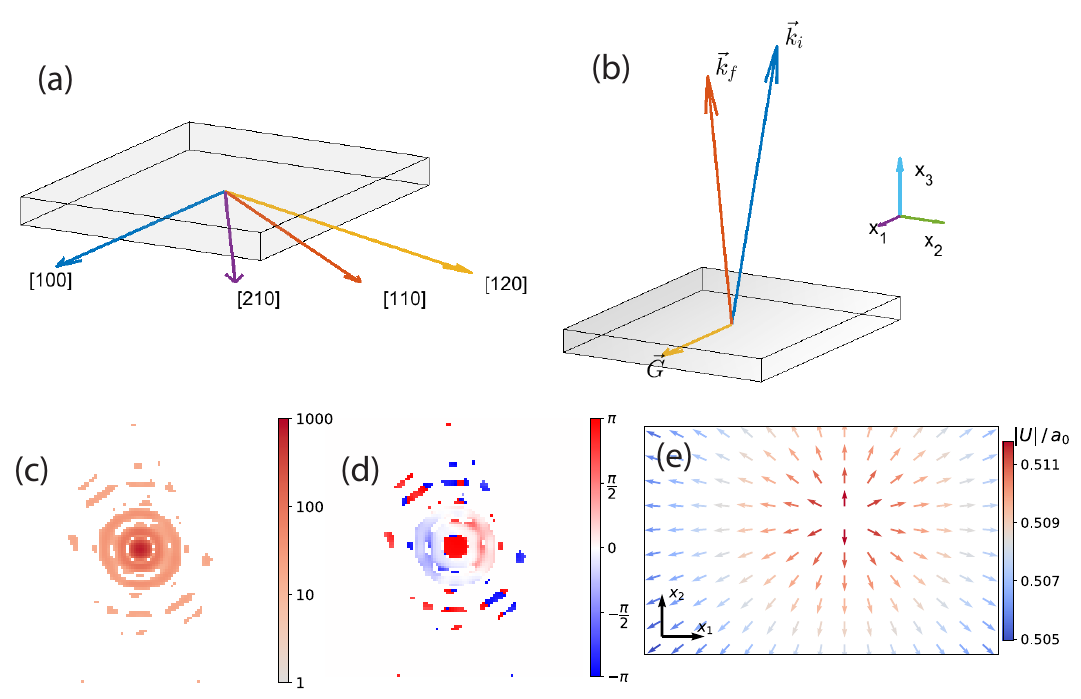}
	\caption{Illustration of 
		(a) Bragg peaks of interest for the simulated thin crystal film, (b) 
		incident ($\kk_i$) and exit($\kk_f$) directions for the $[100]$ Bragg peak (labelled 
		$\vectG$), (c) magnitude of the probe beam along its propagation direction, (d) phase 
		of the probe beam, and (e) the simulated radially decaying displacement profile 
		($\vectU$), 
		where the direction of the arrows show the direction of the lattice displacement, and 
		the color of the arrows show the magnitude of the displacement.}
	\label{fig:experiment_combined}
\end{figure}

We define a single-crystal sample in the laboratory frame $(x_1, x_2, x_3)$ depicted in~\Cref{fig:experiment_combined}, with the incident beam along the $x_3$-direction. 
We adopt a sample geometry that mimics a crystalline free-standing membrane or a thin film affixed to a substrate with different lattice structure. Such systems are plentiful in the fields of functional materials \cite{guo_nl_2021} and micromechanical systems. Further, we note that in such systems where the membrane/film thickness is of order of \SI{\sim 100}{\nm}, lateral heterogeneities in lattice characteristics over micrometer length scales are often of  interest and through-thickness variations of strain are often minor by comparison. 


The experiment that motivated the membrane model construction and the diffraction geometry considered in this work is described in \cite{takahashi_prb_2013}. In that work, a localized x-ray beam (11.8 keV energy) was oriented at near-normal incidence to a single crystal silicon membrane. By tuning the sample angle off of normal incidence by $15.9^\circ$, a Bragg reflection with $hkl$ indices of 220 from the (110) lattice planes could be measured in a Laue geometry with a diffraction vector $\vectG_{hkl}$ contained within the plane of the membrane.
(Here, $\vectG_{hkl} = \kk_i - \kk_f^{hkl}$ with $\kk_i$ and $\kk_f^{hkl}$ being the wave vectors of the incident and exit x-ray beams, each with $|\kk| = 1/\lambda$ where $\lambda$ is the x-ray wavelength.) 
A Bragg ptychography image of a single component of the lattice displacement field in the two-dimensional plane of the membrane was reconstructed from a ptychography series of 220 Bragg peak intensity patterns. In the work we present here, we conceptualize an extension of the work of Takahashi \etal{} by proposing multiple such measurements at different Bragg peaks with in-plane $\vectG_{hkl}$ vectors ($hkl$ reflections for which $l=0$) and spatially resolving a lattice displacement field $\uu(\rr)$ in terms of components $U_1$ and $U_2$ along the $x_1$ and $x_2$ in-plane spatial coordinates via global optimization. 

As shown in Figure 1, for this work, a 3D membrane sample model with extended dimensions along $x_1, x_2$ and limited extent along $x_3$ was constructed. An inhomogeneous lattice displacement field $\uu(x_1, x_2)$ was defined that was self-similar within the $x_3$ dimension of the membrane. A set of Bragg reflections with in-plane diffraction vectors were chosen with diverse in-plane orientation and different reciprocal space magnitudes. At each $hkl$ Bragg reflection, the sample has a distinct scattering amplitude $\vectchi_{hkl}(\bs{r})$ and can be described as a 3D complex-valued real-space object:
\begin{equation}
	\oo_{hkl} =  \vectchi_{hkl}(\rr) e^{2 \pi i \uu(\rr)^T \vdot \vectG_{hkl}}. \label{eq:effective_object}
\end{equation}
where $\vdot$ denotes a matrix multiplication.
Since the Bragg peaks of interest all have $l=0$, we ignore this Miller index in the subsequent text.
For convenience, we represent the effective real-space object for each Bragg peak as a single column vector $\OO_{hk}$ obtained by concatenating the elements of $\oo_{hk}$. 

In the case of Bragg ptychography from a 3D crystal, the x-ray illumination function (or probe) is also 3D and varies with different Bragg peaks due to the different angles of incidence of the beam on the sample and the different orientations of the exit beam \cite{hruszkewycz_oe_2015}. 
We denote the 3D probe that satisfies each Bragg diffraction condition as $\PP_{hk}$ based on a set of incident sample angles and exit beam orientations consistent with a cubic-structured crystal in a diffractometer equipped with a single rotation axis sample stage and a two-rotation-axis detector geometry typical of nanodiffraction beamlines. 
The 2D beam profile that was used to determine $\PP_{hk}$ was one reconstructed from experimental data of a test pattern illuminated with a nanofocused beam, as described in \cite{hill_nl_2018}.
For convenience, we then construct a diagonal matrix $\diag{\PP_{hk}}$ with the elements of $\PP_{hk}$.  Since the crystal is thin in the $x_3$ dimension and we seek to reconstruct a the 2D displacement field of the membrane, we do not require 
the information generated via an angular scan of the object about each Bragg angle. Therefore, for this experiment, 
the expected intensity at the far-field detector plane for each ptychographic scan position can be written as:
\begin{align}\label{eq:mpbp_fwd_model}
	\vectI_{hk,j} = 
	\norm{\FF\vdot\matR_{hk}\vdot\diag{\PP_{hk}}\vdot\SSn_{hk,j}\vdot\OO_{hk}}^2 
	+ 	\bb_{hk,j}
\end{align}
where $j$ indexes the scan positions (with $1\leq j\leq J$), $\SSn_{hk,j}$ is an operator that shifts the object laterally per scan position, $\bb_{hk,j}$ is the background, and $\matR_{hk}$ is the operator that rotates and interpolates the 3D real-space illuminated object from the 
$\xx_1$\nobreakdash--$\xx_2$\nobreakdash--$\xx_3$ coordinate reference frame to the detector 
reference frame for each Bragg reflection (which contains $\kk_f^{hk}$ as one of the orthonormal coordinate directions), then generates the 2D exit wave along the  $\kk_f^{hk}$ direction. 
The construction of the forward model in \eqref{eq:mpbp_fwd_model} is modeled after the one presented in \cite{hill_nl_2018}, where the operator $\matR$ is also described.
These detector reference frames  are typically different for each Bragg peak and, 
consequently, this interpolation is also different. Furthermore, this interpolation also 
needs to account for the fact that the sample-detector distance and the detector pixellation 
could be  different for these different detector geometries---therefore the interpolated 2D object projection 
could  have different voxel sizes as dictated by the Nyquist sampling condition---as well as 
other necessary coordinate transformations 
\cite{li_jac_2020,maddali_jac_2020}.
Finally, by Fourier transforming ($\FF$) the exit wave thus calculated, we generate multiple sets of 
diffraction patterns per 
Bragg reflection consistent with a sample being rastered in the beam in overlapping steps. 
We denote the measured intensities as 
$\yy_{hk,j}$. 

The forward model that we have developed has two key features. First, 
the \textit{phase} in all the effective objects $\OO_{hk}$ (\Cref{eq:effective_object}) is 
due to the \textit{shared} (among all the Bragg peaks) displacement field, and this displacement field is what we 
aim to reconstruct. Second, the amplitudes (or magnitudes) of the objects for the different peaks differ 
only by a global scaling factor that varies with possible differences in the scattering strength of the sample at different Bragg peaks.

\section{Solving the MPBP problem}
\label{sec:optimization}
To formulate the MPBP reconstruction algorithm, we first need to identify the 
variables we want to solve for, then define the error metric that we aim to minimize. 
For this work, we assume that the probe functions are known, that we obtain the intensity 
diffraction data at $D$ different Bragg peaks,
that the lateral dimensions of interest of the membrane can be represented with $N_{x_1}\times N_{x_2}$ pixels, and that the membrane does not change along the $x_3$ direction.
The variables we want to reconstruct are as follows:
\begin{itemize}
	\item the $2N$ displacement variables represented by the matrix $\vect{U}$ defined 
	row-wise as $\vect{U}_n = \begin{pmatrix}u_{x_1,n},\, u_{x_2,n}\end{pmatrix}$, 
	where each	row $\vect{U}_n$ is the transpose of the displacement vector for the voxel 
	$n$, and where 
	$1\leq n\leq N=N_{x_1}\times N_{x_2}$ indexes the voxels. 
\item the $N$ per-pixel amplitude variables for the Bragg peak chosen as reference. We represent these variables as $\vect{\rchi}$ and note that 
	$\vect{\rchi}\geq 0$ elementwise. 
 We note that any of the $D$ Bragg peaks can serve equally as the reference peak.
\item the $D-1$ global scaling factors, represented by the vector $\vect{\alpha}$ (with 
	$\vect{\alpha} > 0$ elementwise),  that 
	we use to scale the object magnitudes for 
	the remaining Bragg peaks. 
        These scaling factors simultaneously account for differences in scattering power of different Bragg peaks 
        and differences in the measurement between Bragg peak ptychography scans (i.e. choice of exposure time). 
\end{itemize}
We therefore have a total $3N + D -1$ unknowns to solve for. We can write the objects (per Bragg peak) in 
terms of these variables as:
\begin{align}\label{eq:object_bragg_peak}
	\OO_{hk} = \begin{cases}
		\vect{\rchi}e^{i \vectU \vdot \vectG_d} \quad \text{if } d=1,\\
		\alpha_{d-1} \vect{\rchi}e^{2\pi i  \vectU \vdot \vectG_d}\quad \text{otherwise},
	\end{cases}
\end{align}
where $\vectG_d$ denotes the scattering vector for the $d$th Bragg peak. We attempt to 
solve for 
these variables by minimizing the Gaussian error metric,
\begin{align}\label{eq:gaussian}
	f(\vectU, \vect{\rchi}, \vect{\alpha})\coloneqq \frac{1}{2}\sum_{hk,j} 
	\norm{\vectI_{hk,j}^{1/2} - 
		\yy_{hk,j}^{1/2}}^2.
\end{align}
The nonlinear minimization problem is then
\begin{align}\label{eq:minimization}
	\text{find } (\vectUstar, \vect{\rchi}_\star, \vect{\alpha}_\star) \in 
	\argmin_{\vectU\in\Reals,\, \vect{\rchi}\in\Reals_{\geq 0},\, 
		\vect{\alpha}\in\Reals_{>0}}{f}.
\end{align}

\subsection{Ambiguities in the displacements: a challenging reconstruction problem}
\label{subsec:model_mismatch}

Previous works \cite{kandel_oe_2019, hruszkewycz_nm_2017,hill_nl_2018} have shown that the  single-peak Bragg ptychographic phase retrieval problem can be solved using a gradient-descent-based update algorithm. However,  the reconstruction of the crystal lattice displacement field, using either multiple such phase profiles (per Bragg peak) or directly from the diffraction datasets, is challenged by an implicit periodicity that leads to a preponderance of 
sub-optimal \textit{local} minima in the landscape for the error metric 
(\Cref{eq:gaussian}).
This necessitates an adaptation so that gradient-descent-based iterative approaches can
find the true solution. In this section, we first demonstrate how the lattice displacement reconstruction problem is affected by ambiguities, then propose a solution to address this challenge. While we focus this discussion on the MPBP setting, the arguments we develop apply to general Bragg coherent imaging of single-crystal lattice displacements. 

%
\begin{figure}[th]
	\centering
	\includegraphics[width=0.95\linewidth]{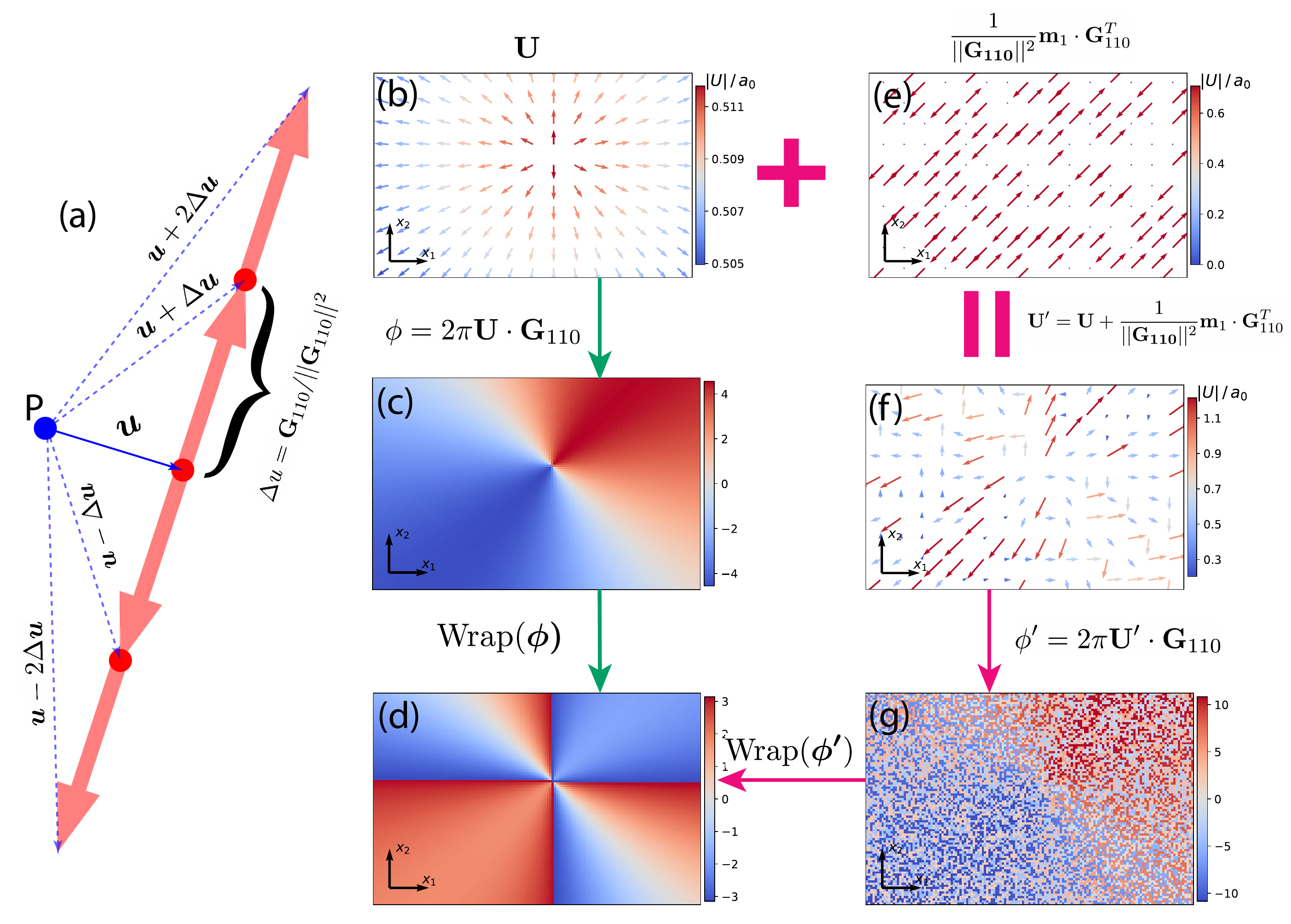}
	\caption{(a) A schematic representation of the ambiguity in the displacement vector at a given pixel. Considering the $[110]$ peak for example, at any location in the reconstruction field of view $P$, the displacements $\uu + m\Delta\uu$ 
	(with $m$ any integer) all give rise to the same phase value.
        (b) The true displacement 
	profile (scaled by the lattice distance $a_0$) is associated with the unwrapped phase profile (c), which after wrapping gives the 
	profile in (d). If we choose $m\in{-1,0,1}$ randomly for every point in the film, we get 
	the field in (e). Adding the fields in (b) and (e), we get the unphysical displacement 
	field (f) that is associated with the phase profile (g). After wrapping, the phase profile 
	in (g) also gives the profile in (d). We therefore cannot distinguish between the 
	displacement fields (b) and (f) based on the object phase for the $\GG_{110}$ peak alone.}
	\label{fig:displacements_periodicity}
\end{figure}

We first consider the effective object for a single Bragg peak $\vectG_1$, which can be written using several equivalent forms,
\begin{subequations}\label{eq:object_periodic}
	\begin{align}
		\OO_1
		&= \abs{\OO_1} e^{2\pi i \vectUstar \vdot \vectG_1}
		\label{eq:object_periodic_1}\\
		& = \abs{\OO_1} e^{2\pi i (\vectUstar \vdot \vectG_1 + \vectm_1)} 
		&&(\text{for any }	\vectm_1\in\mathbb{Z}^{N})
		\label{eq:object_periodic_2}\\
		& =\abs{\OO_1} e^{2\pi i \left(\vectUstar + 
			\frac{1}{\norm{\vectG_1}^2} \vectm_1 \vdot \vectG_1^T\right)\vdot \vectG_1} 
		\label{eq:object_periodic_3},
	\end{align}
\end{subequations}
that are related to one another due to the inherent periodicity in the phases. In fact, the term $\vectU' = \left(\vectUstar + \frac{1}{\norm{\vectG_1}^2} \vectm_1 \vdot \vectG_1^T\right)$ 
in \Cref{eq:object_periodic_3} represents a family of possible lattice displacements that are indistinguishable with respect to $\OO_1$. As a result, any choice of $\vectm_1$ gives rise to the same diffraction patterns as the true displacement $\vectUstar$. 
As an example, we can look at the displacement field corresponding to 
\begin{align}\label{eq:ambiguity_examples}
	\vectm_1 = \begin{pmatrix}1,\,1,\,\dots,\,1\end{pmatrix}^T
	\quad\text{or}\quad
	\vectm_1' = \begin{pmatrix}5,\,-5,\,5,\,-5,\dots\end{pmatrix}^T.
\end{align}
The true displacement field $\vectU_\star$ has $\vectm_1 = (0,0,\dots,0)$, yet the $\vectU$ fields arising from $\vectm_1$ and $\vectm_1'$ both produce the same diffraction patterns. Moreover, $\vectm_1'$ is associated with a highly unphysical spatially discontinuous strain tensor. Consequently, we cannot distinguish the true displacement $\vectUstar$ from any $\vectU'$ using data along just the  Bragg peak $\vectG_1$, even when $\vectUstar \vdot \vectG_1 = \vect{1}$ so that the projection does not cause any loss of information. We illustrate this challenge in ~\Cref{fig:displacements_periodicity}:  ~\Cref{fig:displacements_periodicity}a schematically depicts the family of displacements $\vect{u} \pm m \Delta \tfrac{\vect{G}_{110}}{\norm{\vect{G}_{110}}^2}$ that would all lead to the same  $\vectG_{110}$ phase for any pixel $P$, and \Cref{fig:displacements_periodicity}b-g show the effect of this per-pixel ambiguity for the 2D thin film lattice displacement profile of the membrane model under consideration.

Next,  we examine the case in which we have diffraction data generated along two Bragg peaks 
$\vectG_1$ (for $(h_1, k_1, 0)$) and $\vectG_2$ (for $(h_2, k_2, 0)$). If $\vectG_1^T \vdot \vectG_2=0$, then $\vectU' \cdot \vectG_2 = \vectUstar \cdot\vectG_2$, which means we again cannot isolate the summand $\vectUstar$ using just the diffraction data along these two Bragg peaks. When $\vectG_2^T \cdot \vectG_1 \neq 0$, the additional Bragg peak may reduce the set
of ambiguities that are inherent to the single Bragg peak case. However, to our
best knowledge, there is no formal proof that any combination of noise-free
measurements about specific Bragg peaks provides an unambiguous solution
for the problem. It is also likely that local minima exist for the minimization
problem \Cref{eq:minimization}. Since gradient-based methods can only guarantee converge to local minima, applying these to the optimization problem in \Cref{eq:minimization} is prone to stagnation at these minima --- we observe this in practice in the 4-peak MPBP simulation that we present subsequently.

Our analysis so far shows that the inherent
ambiguities in the displacements lead to spurious global or local minima in the landscape for the error metric $f$ (\Cref{eq:gaussian}, and we propose the following heuristic solution to this challenge.  
 We formulate our solution by noting that, in the MPBP forward model, we first project the pixelwise displacements $\vectU_n$ through the operation $\vectphi (\vectU_n, \vectG_d) = 2\pi \vectU_n\vdot\vectG_d$ per Bragg peak $\vectG_d$.
If we ignore any physics-based 
constraints, the elements of $\vectU$ can attain any value in $(-\infty,\infty)$, and the 
range of 
$\vectphi (\vectU_n, \vectG_d)$ is also $(-\infty,\infty)$. 
In contrast, when we 
invert this model, we calculate these projection through the function $\vectphi_1^\text{inv} (\OO_{1,n}) = 
\arctan\left(\Img{\OO_{1,n}} / \Rel{\OO_{1,n}}\right)$, the range of which is only $(-\pi, 
\pi)$. The gradient-based inversion program does not correct for this mismatch, leading to possibly incorrect $\vectU_n$ values. A natural solution to this mismatch would then be to somehow 
change the range of $\vectphi_1^\text{inv} (\OO_{1,n})$ to $(-\infty,\infty)$: we can achieve this 
through the phase unwrapping procedure \cite{herraez_ao_2002}. Since the phase 
unwrapping procedure removes the $2\pi$ jumps in the phase map, 
they change the range of the phase map from $(-\pi, \pi)$ to $(-\infty, \infty)$. As such, incorporating the phase unwrapping procedure within the MPBP optimization algorithm should be sufficient to correctly solve for the crystal displacements, and we demonstrate this with the subsequent numerical experiments.

\subsection{CDI-based strain reconstruction methods in the literature}
\label{subsec:existing_algos}

While we are not aware of any existing literature that specifically addresses the MPBP 
problem, there exist two primary classes of algorithms that reconstruct crystalline 
displacement (or strain) by combining the diffraction data obtained from multi-peak BCDI 
experiments: the classical \textit{two-step} reconstruction methods (which are the methods of 
choice in the literature), and the newly developed \textit{concurrent} reconstruction methods.

In the two-step reconstruction approach, one
first reconstructs the full effective objects for 
the individual Bragg peaks, then uses the reconstructed phase profiles to 
obtain the displacement and strain profiles 
\cite{robinson_nm_2009,newton_nm_2010,stangl_2013,hofmann_prm_2020}. This 
approach suffers 
from a number of technical challenges. First, it requires the initial solution of a total of 
$2ND$ 
magnitude and phase variables, which can be much higher than the $3N + D -1$ variables that we 
actually require, and is therefore a harder inversion problem. If we have insufficient or 
noisy data along one of the Bragg peaks, then the 
corresponding object reconstruction can contain artifacts or fail altogether, and thereby
significantly deteriorate our desired displacement and strain reconstructions. Second, while  
individual BCDI reconstructions are agnostic to translations of the real-space objects, we 
need to ensure that these individual real-space objects coincide to perform the displacement 
reconstructions, and aligning these can itself be a challenging problem 
\cite{newton_prb_2020}. 

In the concurrent reconstruction approach, we treat the diffraction data generated from 
multiple Bragg peaks as one dataset and use this dataset to simultaneously
reconstruct the real-space objects and the displacement fields. While the existing works that 
use this approach \cite{newton_prb_2020,gao_prb_2021,wilkin_prb_2021} do not specifically solve for the 
variables we cataloged earlier in this section (\viz{} the displacement, magnitude, and 
scaling variables), and solve for individual real-space objects instead, they share the 
information between these objects by 
also tracking the displacement profiles and the real-space support, thereby implicitly solving 
for the same $3N + D -1$ variables instead of the full $2ND$ object variables.

The problem of reconstructing displacement fields that result in phase wraps has been handled in different ways in BCDI literature.
In the two-step reconstruction approach, these phase maps have been first 
unwrapped before the displacement calculations, but, even so, the displacements so calculated 
are themselves often error-prone \cite{ulvestad_sr_2017}. In a recent work, Hofmann \etal{} 
\cite{hofmann_prm_2020} refine the basic two-step approach by using carefully designed phase 
offsets and working with phase gradients to directly reconstruct the strain tensors. On the 
one hand, this method is agnostic to any phase wrap issues. On the other hand, it solves for 
the much larger set of $2ND$ real-space object variables, introduces even more variables in 
the strain reconstruction step, and also requires a careful alignment of the real-space object 
reconstructions. Finally, Maddali \etal{} \cite{maddali_npj_compmat_2023} have recently developed a concurrent reconstruction approach that uses a median filter to minimize any distortions in the displacement field, but this approach may not resolve phase-wrap-associated artifacts at longer length scales and may also deteriorate the spatial resolution of the reconstruction. As such, a concurrent reconstruction approach that robustly addresses the model 
mismatch problem very desirable. 

\subsection{The MPBP reconstruction algorithm}

In this section, we outline a MPBP reconstruction algorithm that, in contrast to the 
approaches described above (\Cref{subsec:existing_algos}), 
directly solves for the desired displacement 
($\vectU$), 
magnitude ($\vectchi$), and scaling ($\vectalpha$) variables, in a manner analogous to that developed in \cite{maddali_npj_compmat_2023}, but for the case considered here of ptychography data measured at multiple Bragg conditions, with specific adaptations aimed at the phase wrap problem. 

To begin with, we note that 
magnitude and scaling variables should be constrained so that the $\vectchi \geq 0$ and 
$\vectalpha >0$; ignoring these constraints in the optimization can lead to stagnation in the 
minimization.  
While it is certainly possible to apply these constraints (through projection steps, for 
example), we  implement this constraint indirectly by defining the auxiliary 
variables $\tilde{\vectchi}\in\Reals^N$ and $\tilde{\vectalpha}\in\Reals^{D-1}$ that satisfy 
the relations
\begin{align}\label{eq:aux_vars}
	\vectchi = e^{\tilde{\vectchi}}, \quad\text{and}\quad \vectalpha = e^{\tilde{\vectalpha}}.
\end{align}
Solving for these auxiliary variables naturally enforces the constraints, and we can 
use a small threshold at the end of the optimization procedure to set $\vectchi=0$ in regions not illuminated by the probe.

We now define a $2\times D$ matrix $\vect{H}$ columnwise as $\vect{H}_{\bullet,d} = 
\begin{pmatrix}\vectG_{d}\vdot \vecte_1\\ \vectG_d\vdot\vecte_2\end{pmatrix}$, where 
$\vecte_1$ 
and 
$\vecte_2$ are the unit vectors along the $\xx_1$ and $\xx_2$ directions respectively. We can 
then simultaneously calculate the phases for all the real-space objects as
\begin{align}\label{eq:phase_displacement_matrix}
	\vect{\Phi} = \vectU\vdot \vect{H},\quad\text{and, conversely,}\quad 
	\vectU = \vect{\Phi}\vdot \vect{H}^T \vdot \left(\vect{H}\vdot\vect{H}^T\right)^{-1},
\end{align} 
where $\vect{\Phi}$ is a $N\times D$ matrix whose columns contain the phase maps for the 
individual Bragg peaks, and where the second expression comes about because $\vect{H}$ is a 
full-rank rectangular matrix. If we define a phase wrapping operation $\mathcal{W}$ that simultaneously 
calculates the wrapped phase for all the real-space projections of the displacement field, and we define the phase 
unwrapping operation $\mathcal{W}^{-1}$ that simultaneously calculates the unwrapped phase for 
all the real-space objects, we can now define a projector:
\begin{align}\label{eq:unwrap_projection}
	\Pi_U (\vectU)= \mathcal{W}^{-1}\{\mathcal{W}\{\vectU\vdot \vect{H}\}\}\vdot \vect{H}^T 
	\vdot 
	\left(\vect{H}\vdot\vect{H}^T\right)^{-1}.
\end{align}
Under ideal circumstances, this projector takes the displacement field, which could contain 
the periodicity-induced discontinuities, calculates the associated phase maps, removes the 
discontinuities via the phase unwrapping procedure, then calculates the now artifact-free 
displacement fields, thereby addressing the minimization challenges we discussed in 
\Cref{subsec:model_mismatch}. We note that this is a heuristic operation that depends 
greatly on the properties of the chosen phase unwrapping algorithm, and could be error-prone 
in practical applications. However, we find that 
this projector, used within the reconstruction algorithm outlined below (\Cref{alg:mpbp}), 
works quite 
well in our numerical simulations. 
\begin{algorithm}[H]
	\caption{Single iteration of the minibatch MPBP reconstruction algorithm}
	\singlespacesmall
	\begin{algorithmic}[1]
		\Require Initial guesses $\vectU^0$, $\tilde{\vectchi}^0$, $\tilde{\vectalpha}^0$.
		\Require Minibatch size $b$.
		\State Randomly shuffle the $JD$ diffraction patterns irrespective of the $J$ probe 
		positions (per peak) and the $D$ Bragg peaks. 
\begin{align*}
			\mathfrak{J} =\{1,2,...,JD\}\xrightarrow{\text{shuffle}}\mathfrak{J}'
		\end{align*}
\For {$t=1$ to $JD/b$}
		\State With the diffraction patterns indexed as $j\in\mathfrak{J}'$, calculate the 
		partial derivatives:
\begin{subequations}\label{eq:minibatch_grads}
			\begin{align}
				\pgrad_U f^t &\coloneqq \sum_{j=b(t-1)+1}^{bt} \pgrad_U f(\vectU^{t-1}_j),\\
\pgrad_{\tilde{\rchi}} f^t &\coloneqq  \sum_{j=b(t-1)+1}^{bt} 
				\pgrad_{\tilde{\rchi}} 
				f(\tilde{\vectchi}^{t-1}_j),\quad\text{and}\\
\pgrad_{\tilde{\alpha}} f^t &\coloneqq  \sum_{j=b(t-1)+1}^{bt} 
				\pgrad_{\tilde{\alpha}} 
				f(\tilde{\vectalpha}^{t-1}_j).
			\end{align}
		\end{subequations}
\State Use these partial derivatives within the Adam algorithm \cite{kingma_corr_2014} 
		to 		calculate $\vectU^t$, $\tilde{\vectchi}^t$, and $\tilde{\vectalpha}^t$.
		\State Set $\vectU^t\longrightarrow \Pi_U(\vectU^t)$.
\EndFor
	\end{algorithmic}
	\label{alg:mpbp}
\end{algorithm}

We present the steps taken within a single iteration of our MPBP reconstruction approach in 
\Cref{alg:mpbp} and note that this is a variation of the basic minibatch ptychography 
algorithm presented in \cite{kandel_oe_2019}[Algorithm 1]. However, the use of the $\vectU$, 
$\tilde{\vectchi}$, and $\tilde{\vectalpha}$ variables, instead of the $(\Rel{\OO}, 
\Img{\OO})$ variables in \cite{kandel_oe_2019}, for the gradient calculations 
requires 
extra coordinate transformations and other mathematical manipulation. Indeed, this adds to the 
already significant algebra that would be required to calculate the gradients just for the 
basic MPBP model. Therefore, by using the AD framework (which is implicit in 
Line 3 of  
\Cref{alg:mpbp}), we not only avoid such tedious and error-prone algebra, but also 
take advantage of the state-of-the-art parallelized implementations of the mathematical 
operations (including the complex interpolations) required for both the MPBP forward model and 
gradient calculations\cite{kandel_oe_2019,maddali_npj_compmat_2023,kandel_oe_2021,du_oe_2021}.

\section{Numerical experiment}
\label{sec:numeric}
To test our proposed reconstruction algorithm, we use a numerical experiment with the basic 
experimental parameters listed in \Cref{tab:sim_details}, wherein we simulate a thin film 
with a cubic lattice structure, which has a radially decaying 2D lattice distortion field emanating from a point in the field of view. 
This would be consistent with the displacement field associated with a local stress concentration in a crystal membrane, for example.
For convenience, we define an orthonormal ``lab frame'' with 
the axes $(\xx_1, \xx_2, \xx_3)$ and with the voxel size $5.84 \times 5.84 \times 5.84$ cubic microns, in which 
the simulated membrane is centered at the origin, and is aligned with its surface along the \plane{\xx_1}{\xx_2} so 
that it has the dimensions of $137 \times 85\times 17$ voxels.
As discussed in \cite{maddali_npj_compmat_2023}, the voxellation in the lab 
frame needs to be fine enough to obtain faithful rendering of the object after  
interpolation and projection to the detector frames of each Bragg peak. 

To create a simulated data set from this membrane, we perform numerical ptychography experiments at each of the $[100]$, $[110]$, $[120]$, and $[210]$ Bragg peaks for the cubic lattice. 
(In an experiment setting, the choice of available peaks is dictated by the symmetry of the underlying crystal.) 
The \textit{magnitudes} of 
the real-space objects ($|\OO_{hk}|$) are 
scaled from unity to \numlist{0.004; 0.0035; 0.0021; 0.0010} for the $[100]$, $[110]$, $[120]$, and 
$[210]$ peaks respectively. 
This set of scaling parameters was chosen arbitrarily and has the effect of creating differing levels of low count rate diffraction intensity patters that yield appreciable shot noise when used to generate consistent Poisson counting statistics, as was done here. 
Using different scaling parameters also necessitates that the relative scaling factors be reconciled by the reconstruction algorithm via the variable $\vectalpha$.
We show the orientations of the chosen  $\vectG_{hk}$ vectors and the radial profile of simulated lattice displacement field in 
\Cref{fig:experiment_combined}, the individual components of the displacement in 
\Cref{fig:displacements}, the phase maps along each of the Bragg peaks in 
\Cref{fig:phases}, and the magnitude maps in \Cref{fig:magnitudes}.

\begin{table}[H]	
	\caption{Experimental details}
	\label{tab:sim_details}
	\centering
	\begin{tabular}{|c|c|}
		\hline
		\textbf{Parameter} & \textbf{Value} \\
		\hline
		Wavelength & \SI{1.377}{\Angstrom} (\SI{9}{\kilo\eV}) \\
		\hline		
		Detector &  $150 \times 150$ pixels \\
		\hline
		Detector pixel size & \SI{55}{\micro\meter} \\
		\hline
		Sample detector distance & \SI{0.35}{\meter} \\
		\hline
		Film length ($\xx_1$) & \SI{0.8}{\micro\meter}\\
		\hline
		Film width ($\xx_2$) & \SI{0.5}{\micro\meter}\\
		\hline
		Film thickness ($\xx_3$) & \SI{0.1}{\micro\meter}\\
		\hline
		Lattice distance (cubic) & \SI{3.905}{\Angstrom}\\
		\hline
		Numerical window ($\xx_1$, $\xx_2$, $\xx_3$) & $200	\times 200\times 100$ voxels\\
		\hline
	\end{tabular}
\end{table}

In our ptychography simulations, we use a 2D probe profile experimentally measured from a Fresnel zone plate x-ray focusing optic at the HXN beamline at NSLS-II \cite{chu_bnl_2015}, of size 
$100\times 100$ pixels at focus (with each pixel of size 
\SI{1}{\nano\meter}). 
We threshold to zero any pixels in the probe below an intensity cutoff of $2\%$ of the maximum intensity to reduce aliasing in the simulated diffraction patterns. 
Using the HXN diffractometer geometry, we calculated the set of 
sample incident angles and detector position exit angles that satisfy the Laue condition for 
each of the Bragg peaks of interest. 
For each peak, we interpolate the probe to produce the 3D 
illumination volume at the appropriate incident angle, with the beam profile set to be 
constant along the propagation direction through the numerical window. We then scan the thin 
film using a $9\times 9$ grid, with a step size of \SI{6}{pixels}, along the 
\plane{\xx_1}{\xx_2}. 
At every beam position, a diffraction pattern is generated by applying the forward model in \eqref{eq:mpbp_fwd_model}, the scaling values discussed above were applied, and Poisson noise was added. 
The mean count rates of the diffraction patterns at each Bragg peak were $\sim$ 83, 65, 52, and 12 photons per detector pixel respectively. 

During the reconstruction, we solved for an $\xx_1$\nobreakdash--$\xx_2$ area of 
$147 \times 95$ pixels, stacked appropriately to produce the 3D membrane structure, 
initialized as an 
array of zeros, and placed centrally in the numerical window. The film support along the 
$\xx_1$ and $\xx_2$ directions are set to be \SI{5}{pixels} larger than the actual film 
dimensions to allow for limited resolution effects due to the low 
photon count \cite{chamard_sr_2015}. 
We performed reconstructions of the simulated data set using the following reconstruction 
approaches in order to enable a direct comparison of results:
\begin{itemize}
	\item \textbf{TS}: the two-step approach (\Cref{subsec:existing_algos}), 
	where we first solve for the individual real-space objects (for each Bragg peak), unwrap 
	the phases calculated for these objects, then calculate the displacements from the 
	unwrapped phases. We use the AD-based approach described in \Cref{alg:mpbp}, but with the 
	$4N$ phase variables (denoted as $\vect{\phi}_{hk}$) instead of the $2N$ displacement 
	variables, and with the initial Adam update step sizes set to \numlist{0.01; 0.1; 0.1} for 
	the $\vect{\phi}_{hk}$, $\tilde{\vectchi}$, and $\tilde{\vectalpha}$ variables 
	respectively. We thereby solve for each of the real-space objects while utilizing the 
	shared magnitude information between the objects.
\item \textbf{C-W}: the concurrent-wrapped approach that uses the basic algorithm in 
	\Cref{alg:mpbp} but without any phase unwrapping, \ie{} without the projection step (in 
	Line 5), and with the Adam update sizes set to  \numlist{0.01; 0.1; 0.1} 
	for the $\vectU$, $\tilde{\vectchi}$, and $\tilde{\vectalpha}$ variables 
	respectively. In this test, no additional strategies such as local smoothing filters aimed to mitigate phase wrap discontinuities were implemented to represent baseline performance of this algorithm.
\item \textbf{C-UW}: the concurrent-unwrapped approach outlined in \Cref{alg:mpbp}. For 
	the phase unwrapping step, we use the algorithm \cite{herraez_ao_2002} implemented within 
	the ``scikit-image'' platform  \cite{vanderwalt_peerj_2014}. The Adam update sizes are set 
	to  \numlist{0.01; 0.1; 0.1} 
	for the $\vectU$, $\tilde{\vectchi}$, and $\tilde{\vectalpha}$ variables 
	respectively. 
\end{itemize}
All the reconstructions used a minibatch size of \num{50} diffraction patterns, for a total of 
\num{1000} iterations of reconstruction. The update step sizes and the minibatch sizes were 
obtained through the trial and error approach outlined in \cite{kandel_oe_2019}. The 
calculations were performed using the Tensorflow AD platform \cite{abadi_corr_2016}.

\section{Results}
\label{sec:results}
\begin{figure}[th]
	\centering
	\includegraphics[width=0.9\linewidth]{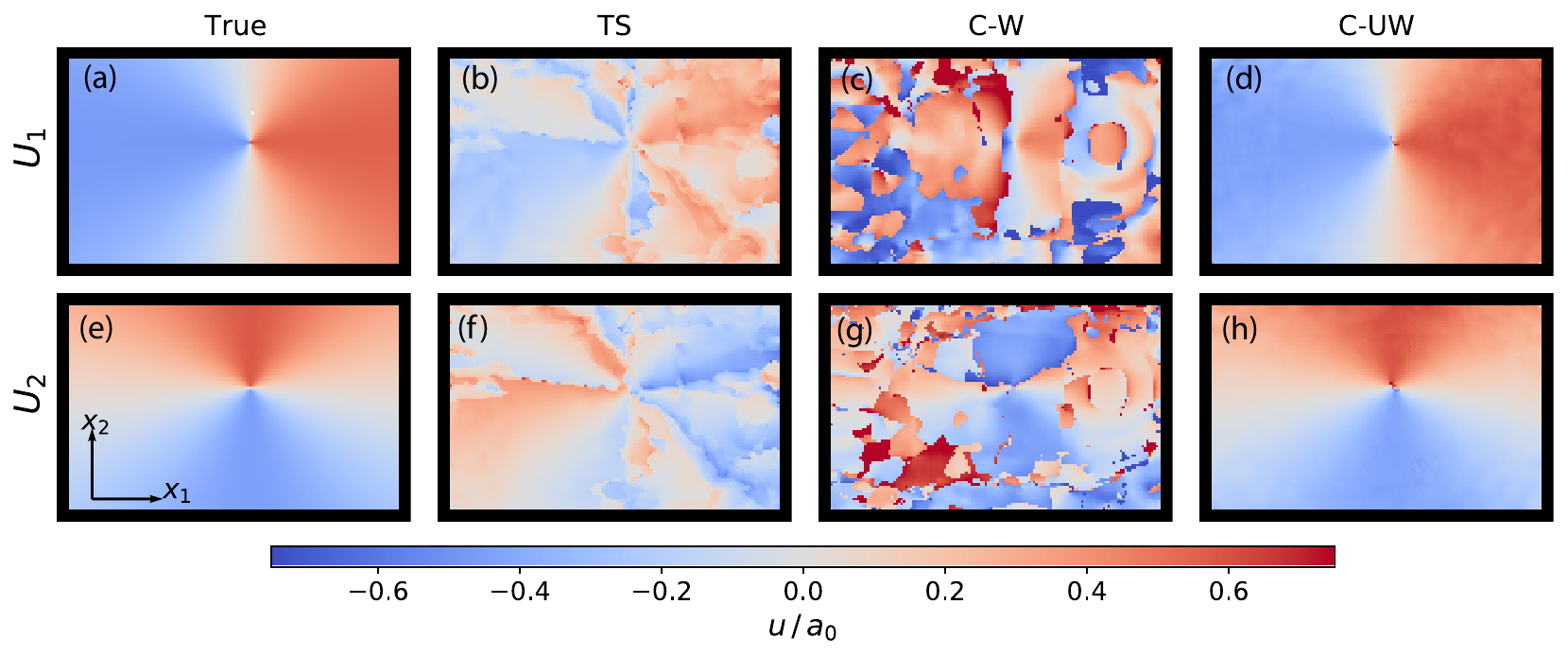}
	\caption[Displacement reconstructions for the MPBP experiment.]{Reconstructed displacement 
		components. (a,e) Simulated (true) values. The 
		reconstructions obtained through the (b,f) two-step (TS) reconstruction method and the 
		(c,g) concurrent method without the phase unwrapping step (C-W) show significant 
		artifacts. The (d,h) concurrent-unwrapped (C-UW) reconstructions show close agreement 
		with 
		the true values.}
	\label{fig:displacements}
\end{figure}
\begin{figure}[th]
	\centering
	\includegraphics[width=0.9\linewidth]{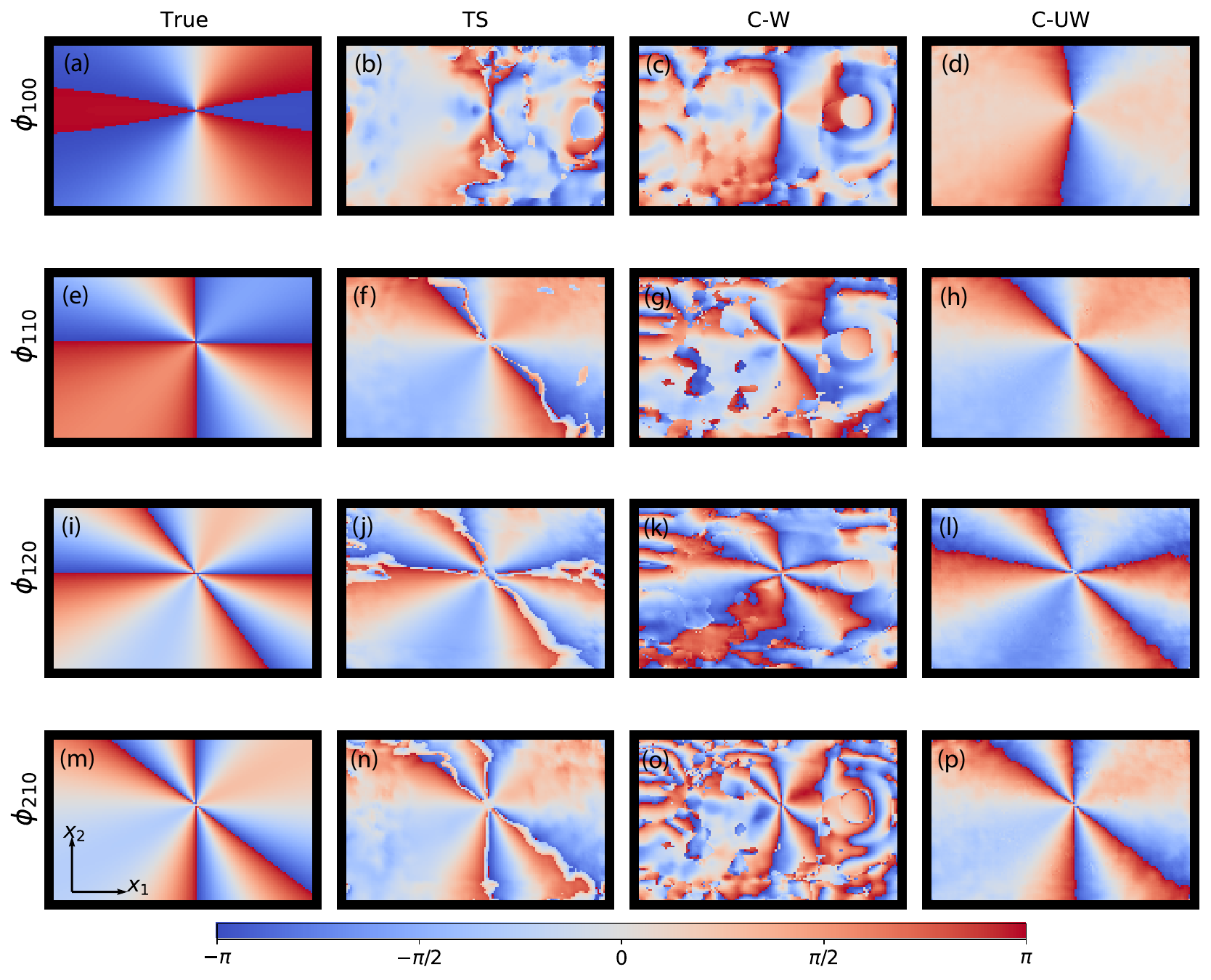}
	\caption[Phase reconstructions for the MPBP experiment.]{Reconstructed phase maps for each 
		Bragg peak. (a,e,i,m) Simulated (true) values. 
		The (b,f,j,n) TS phase maps and the (c,g,k,o) C-W phase maps show significant 
		artifacts 
		primarily at locations coinciding with discontinuous phase jumps (for both the TS and 
		C-W 
		cases), and other unrelated locations (for the C-W case). The C-UW 
		phase maps agree closely with the true values.}
	\label{fig:phases}
\end{figure}
\begin{figure}[th]
	\centering
	\includegraphics[width=0.9\linewidth]{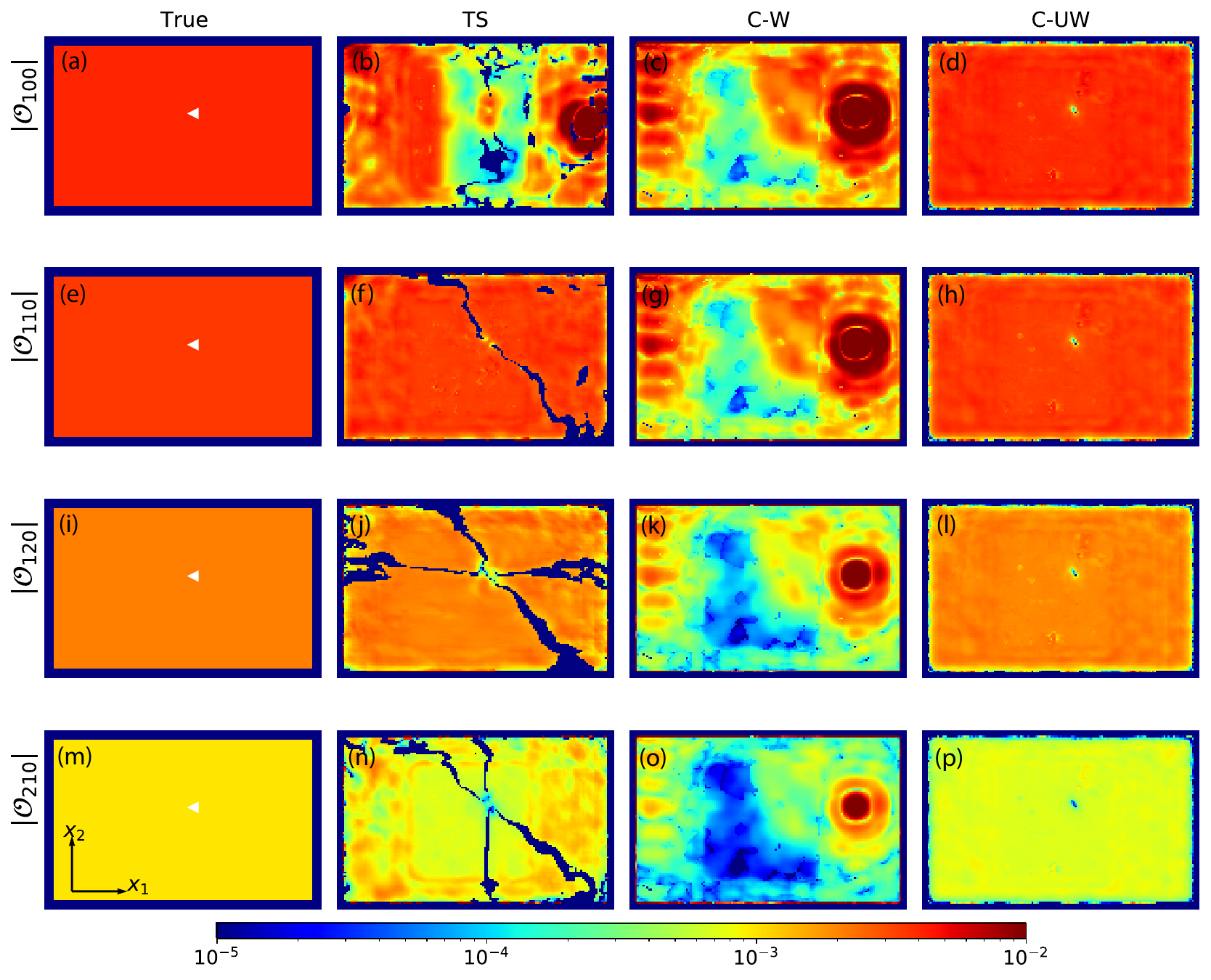}
	\caption[Magnitude reconstructions for the MPBP experiment.]{Reconstructed magnitude maps 
		for each Bragg peak. (a,e,i,m) Simulated (true) 
		values, where the white triangle indicates the location of the point distortion in the 
		lattice displacements. The (b,f,j,n) TS magnitude maps, which are reconstructed 
		individually for each Bragg peak, are inaccurate primarily at locations coinciding 
		with discontinuous phase jumps. Among the (c,g,k,o) C-W and (d,h,l,p) C-UW methods, 
		which 
		both reconstruct a reduced number of magnitude variables, the C-W results contain 
		major 
		artifacts while the C-UW results maps agree closely with the true values.}
	\label{fig:magnitudes}
\end{figure}

In \Cref{fig:displacements,fig:phases,fig:magnitudes} we present the reconstructions 
obtained for the lattice displacement, the phase maps, and the magnitude maps respectively. 
First, we examine the TS case, where \Cref{fig:phases,fig:magnitudes} show that the 
individual real-space object reconstructions are highly inaccurate, with artifacts for both 
phase and magnitude maps that are most pronounced in 
regions with phase discontinuities.
The locations of these phase discontinuities are 
different for the real-space objects from each of the Bragg conditions. In these regions, the magnitude maps show "gaps" (regions of low 
magnitude)---a reconstruction 
artifact that has been observed previously in 
the literature \cite{gao_prb_2021,huang_prb_2011,newton_prb_2010}.  As a 
consequence, the 
displacement calculated  
from the real-space phase maps are also discontinuous and inaccurate, as shown in \Cref{fig:displacements}(b,f).

Next, we can see that the C-W reconstructions, where we directly reconstructed the 
displacement and magnitude maps without accounting for phase wrapping, show significant 
discontinuities in the displacement reconstructions (\Cref{fig:displacements}(c,g)). Similarly, the phase maps (\Cref{fig:phases}(c,g,k,o)) and the reconstructed amplitudes projected to the $\vectG_{hk}$ vectors (\Cref{fig:magnitudes}(c,g,k,o)) both show prominent artifacts.
The similarity of the artifacts in the magnitudes of the C-W reconstruction arise because in this approach the magnitude is scaled and shared among all Bragg peaks. 

Finally, we can see that C-UW reconstructions algorithm accurately reproduces the simulated 
displacement and magnitude maps. In particular, while the magnitude maps 
(\Cref{fig:magnitudes}) still contain a gap in the location of the stress concentration,  
they do not have either the strong phase artifacts (seen in the TS 
reconstructions) or the circular artifact (seen in the C-W reconstructions). The reconstructed 
displacement maps, and consequently the phase maps, are artifact-free.

 
\section{Discussion and conclusion}
\label{sec:discussion}
In summary, we propose a multi-peak Bragg ptychography reconstruction framework to combine 
diffraction ptychographic datasets generated at multiple Bragg reflections to image the full 
lattice displacement and strain profiles. 
The results in \Cref{fig:displacements,fig:phases,fig:magnitudes} demonstrate 
that our proposed approach can reconstruct both the displacement and the 
real-space magnitude profiles measured in the MPBP experiment. Compared to the TS procedure, 
this algorithm: i) significantly reduces the number of variables of interest, which makes for 
a more robust reconstruction, and ii) efficiently utilizes diffraction datasets that could be 
insufficient for individual real-space object reconstructions. By incorporating the phase 
unwrapping step, this algorithm also addresses the periodicity-driven model mismatch in the 
MPBP problem. 

To apply the proposed algorithm to general experiment datasets, we need to be mindful about a 
few different aspects of the algorithm. First, we note that phase unwrapping is itself a 
difficult problem, particularly for objects with steep phase changes. Moreover, while our 2D 
phase unwrapping problem can be solved efficiently \cite{herraez_ao_2002}, its 3D extension 
can be significantly more time consuming \cite{abdul_isop_2005,cusack_neuroimage_2002} and 
difficult to solve. In this context, it could be preferable to replace the phase unwrapping 
procedure altogether 
with a direct  physics-based solution to the periodicity-driven model mismatch problem.
Second, the reconstruction could contain unexpected artifacts if the datasets along the different peaks 
have significantly different photon counts (as we observe for the C-W case in the current 
work), in which case reweighing the diffraction datasets could produce more optimal results. 
Third, registry of the fields of view scanned at the different Bragg angles is imperitive for this approach and has been realized in Bragg scanning nanodiffraction experiments by use of fluorescing fiduciary marks intentionally deposited on the sample surface. 
Finally, the reconstruction procedure is sensitive to the choice of the optimization 
hyperparameters, which therefore require careful tuning. We hope to address this in the 
future by designing more robust second order minibatch optimization procedures \cite{kandel_oe_2021}.
 


Since the proposed algorithm is simple in concept, it can be either 
incorporated within existing concurrent multi-peak BCDI workflows, or extended (via the AD framework) for 
applications in general BCDI or Bragg ptychography experimental models. We can also envision 
straightforward applications of the proposed method to simpler experimental models, such as 
the recently demonstrated ``simultaneous'' multi-Bragg peak CDI experiments 
\cite{lauraux_crystals_2021}. In 
general, we expect that our algorithm marks a concrete step in 
connecting the ever-improving coherence of synchrotron sources to characterization of 
functional materials under real-world working conditions.

\section{Acknowledgements}

This work (concept development, forward model development, algorithm design) was supported by the U.S. Department of Energy (DOE), Office of Science, Basic Energy Sciences, Materials Science and Engineering Division.
Algorithm testing was supported by the Laboratory Directed Research and Development (LDRD) funding from Argonne National Laboratory, provided by the Director, Office of Science, of the U.S. Department of Energy under Contract No. DE-AC02-06CH11357. 
We gratefully acknowledge the computing resources provided on Swing, a high-performance computing cluster operated by the Laboratory Computing Resource Center at Argonne National Laboratory.
This research uses the resources of the Advanced Photon Source, a U.S. DOE Office of Science User Facility operated for the DOE Office of Science by Argonne National Laboratory under contract No. DE-AC02-06CH11357. 
This research used the 3-ID Hard X-ray Nanoprobe (HXN) beamline  of the National Synchrotron Light Source II, a U.S. Department of Energy (DOE) Office of Science User Facility operated for the DOE Office of Science by Brookhaven National Laboratory under Contract No. DE-SC0012704.

\printbibliography

\end{document}